\documentclass[twocolumn,twoside,slac_two]{revtex4}
\usepackage{graphicx}
\usepackage{fancyhdr}
\usepackage{epstopdf}
\newcommand{\degree}{\ensuremath{^\circ}}

\begin{document}

\title{Multi-wavelength Observations of Markarian 501}

%

\author{D. Gall}
\affiliation{Purdue University, West Lafayette, IN 47907, USA}
\author{for the VERITAS collaboration}
\affiliation{For full author list, see http://veritas.sao.arizona.edu/conferences/authors?icrc2009}

\begin{abstract}
The very high energy (VHE; E $>$ 100 GeV) $\gamma$-ray blazar Markarian 501 has a history of extreme spectral variability and is an excellent laboratory for studying the physical processes within the jets of active galactic nuclei. A short term multi-wavelength study of Markarian 501 was coordinated in March 2009 using the Suzaku X-ray satellite as well as the VERITAS and MAGIC experiments to cover the critical X-ray and VHE $\gamma$-ray bands. The results of the quiescent-state observations with VERITAS and Suzaku are combined with public data from the Fermi Gamma-ray Space Telescope and compared to historical observations of the source during an extreme outburst, to examine the spectral variability, particularly how the spectral energy distribution varies with flux.
\end{abstract}

\maketitle

\thispagestyle{fancy}


\section{Introduction}

Blazars are a sub-class of active galactic nuclei (AGN), and they are the dominant extra galactic population in $\gamma$-rays. They have been observed to show rapid variability and non-thermal spectra, presenting a broad continuum across nearly the entire electromagnetic spectrum, implying that the observed photons originated within highly relativistic jets oriented very close to the observerÕs line of sight \cite{Urry}. This orientation results in Doppler beaming that boosts the intensity and frequency of the observed jet emission.  Therefore, blazars make excellent laboratories for studying the physical processes within the jets of AGN. They were among the first sources to be detected in the VHE band and today there are 28 known VHE $\gamma$-ray blazars\footnote{http://tevcat.uchicago.edu/}. 

The spectral energy distribution (SED) of Markarian 501, a VHE $\gamma$-ray blazar, characteristically shows a double peaked structure.  These peaks occur at keV and TeV energies when plotted as $\nu$F$\nu$ vs $\nu$.  This structure is common among all VHE $\gamma$-ray blazars, and several models have been developed to account for the double peaked structure.  

These models attribute the peak at keV energies to synchrotron radiation from relativistic electrons and positrons within the blazar jets.  The models differ in accounting for the source of the TeV peak.  The models are generally divided into two classes:  leptonic and hadronic, named for their attributed source for the TeV peak.  The leptonic models advocate inverse Compton scattering to TeV energies of either the synchrotron photons from within the jet or an external photon field \citep[e.g.,][]{Marscher,Maraschi,Dermer,Sikora}.  The hadronic models, however, account for the TeV emission by $\pi_0$ or charged pion decay with subsequent synchrotron and/or Compton emission from decay products, or synchrotron radiation from ultra-relativistic protons \citep[e.g.,][]{Mannheim, Aharonian, Pohl}.

Observationally, Markarian 501 has been known to undergo both major outbursts on long time scales and rapid flares on short time scales, most prominently at keV and TeV energies. During these outbursts, both of the SED peaks have been observed to shift towards higher energies.  During the most extreme cases, the synchrotron peak has been observed above 100 keV.  Historically, the SED has been measured in the VHE band only during outbursts, due to the lower sensitivity of the previous generations of instruments.  This work attempts to provide state of the art short-term multi-wavelength measurements of the quiescent state of Markarian 501.  These measurements are then compared to an extreme outburst of the source that was observed in 1997 \cite{Pian, Catanese}.

\section{OBSERVATIONS}

\subsection{X-ray}

The Suzaku X-ray observatory, the product of a collaboration between institutions in the United States and Japan, is an excellent tool for studying the broadband spectral energy distribution (SED) of sources such as Markarian 501.  Its multiple instruments provide a broad energy range from 0.2 - 600 keV.  Suzaku has two operating instruments for studying X-ray emission: the X-ray imaging spectrometer \citep[XIS;][]{xis} and the hard X-ray detector \citep[HXD;][]{hxd}.  A third instrument, the X-ray Spectrometer malfunctioned after launch and is no longer operational. 

The XIS instrument consists of four X-ray telescopes with imaging CCD cameras in their focal plane.  Three of the CCDs are front illuminated, and one CCD is back illuminated to provide extended sensitivity to soft X-rays.  The combined energy range of these CCDs is 0.2 - 12.0 keV.  The HXD is a non-imaging instrument that expands the energy sensitivity of Suzaku to the 10 - 600 keV band by using two types of detectors.  Silicon PIN diodes provide sensitivity in the range 10 - 70 keV, and GSO scintillators placed behind the PIN diodes provide sensitivity in the range 40 - 600 keV.  One limitation of Suzaku with respect to attaining simultaneous multi-wavelength observations is that it resides in a low earth orbit, so observations are subject to frequent earth occultations, limiting continuous temporal coverage.

Observations of Markarian 501 were carried out with the Suzaku X-ray satellite from 2009-03-23 UT 18:39 to 2009-03-25 UT 07:59 (sequence number 703046010).  After good time interval selection, a total of approximately 72 ks of live time remained.  The telescope operated with a nominal pointing that optimizes the effective area of the HXD.  Even with this pointing, the maximum useful energy for these observations was around 40 keV.  The HXD/PIN non X-ray background is provided by the HXD team, and the cosmic X-ray background was accounted for by estimating the PIN response to the flat emission distribution.

The XIS observations were requested in 1/4 window mode to reduce the occurrence of pile up.  Quality cuts were made on the data, and XIS events were extracted from a circular region centered on Markarian 501.  Rectangular regions far from the source were chosen for the  background regions.  The auxiliary and response files were created using tools developed by the Suzaku team.  After good time interval selection, a total of approximately 72 ks of live time remained.

%


\begin{figure}[t]
\centering
\includegraphics[width=80mm]{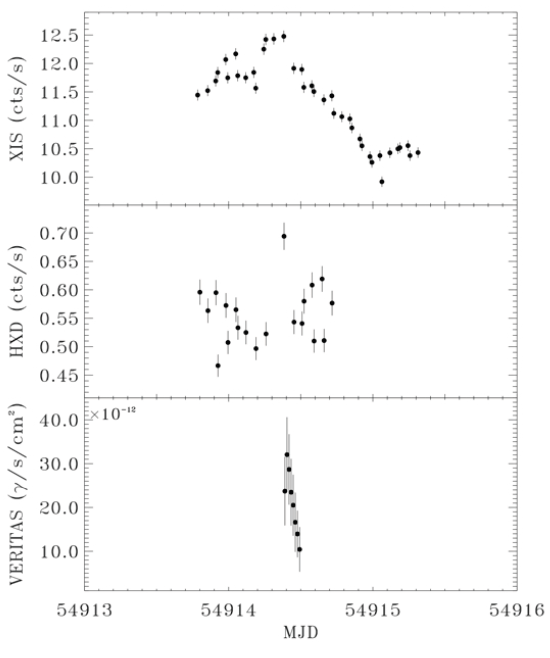}
\caption{Preliminary light curves from Suzaku HXD, XIS and VERITAS.} \label{lc}
\end{figure}

\subsection{VHE $\gamma$-ray}
VERITAS is an array of four 12 m diameter IACTs each focusing light on a camera consisting of 499 PMTs.  Utilizing multiple telescopes provides a stereoscopic view of showers, making it possible to significantly reduce muon events.  This substantially improves the low energy performance of the detector since images produced from nearby muons are otherwise virtually indistinguishable from images produced by low-energy gamma-rays.  Therefore the array allows for a lower energy threshold than possible with a single telescope of the same size operating alone.  In addition, stereoscopic observations facilitate a much improved determination of the core position of showers and thus provide improved energy resolution and background rejection.  The increased sensitivity of VERITAS over previous generations of instruments allows for a high quality spectrum of Markarian 501 to be collected in a relatively short time, even in the quiescent state.

On 2009 March 24 2008, VERITAS took data on Markarian 501 while operating in {\it wobble} mode, with an offset angle of 0.5$\degree$.  To provide background estimations using this observation mode, we follow the reflected-region model \cite{Berge} where one collection region is placed at the source position and others of equal size are placed at equal offsets from the center of the field of view and used for background measurements.  VERITAS observed Markarian 501 for 2.6 hours from 9:11 UT to 11:58 UT using a series of 20 minute runs, with a live time of 2.4 hours.  The data were processed using standard VERITAS analysis cuts, and the total significance for this observation was 9.5 $\sigma$ with an average significance per run of 3.3 $\sigma$.
\begin{figure}[t]
\centering
\includegraphics[width=80mm]{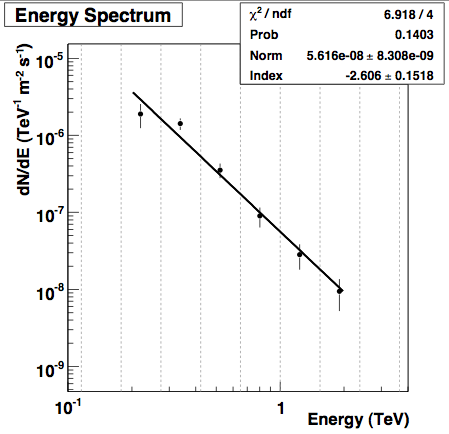}
\caption{Preliminary VHE spectrum from VERITAS, including the best-fit power law.} \label{tevspec}
\end{figure}

\subsection{HE $\gamma$-ray}
The operation of the Fermi Space Telescope allows for measurements to be taken in an energy range that has been out of reach since the Compton Gamma-ray Observatory de-orbited in 2000.  It also has a significant improvement in sensitivity over previous instruments, allowing for spectral points to be measured not only in the flaring state, but also over the course of several days in the quiescent state.   

A preliminary measurement based on public data from the Fermi Gamma-ray Space Telescope was used to include a contemporaneous data point in the broadband SED. There were not enough statistics from the two days coincident with the Suzaku observation to extract spectra, so the preliminary Fermi result shown includes data from March and April 2009 to increase the significance of the spectral point.  The data was analyzed following the guidelines provided by the Fermi team\footnote{http://fermi.gsfc.nasa.gov/ssc/data/analysis/documentation/Cicerone/}.  The disadvantage to the use of a wider time window is the inclusion of large amounts of data that are not simultaneous with the other observations.  At the time of writing of these proceedings, an updated analysis has been provided by the Fermi group using data from only seven days, centered on the Suzaku observation.  This updated SED shows a lower flux in the Fermi energy band, and these updated results will be presented in a forthcoming paper.  

\subsection{Results}

The combined data provide
excellent coverage of the quiescent state of this
extreme VHE blazar.  After following the procedures described above and detailed in the ABC guide for Suzaku data analysis\footnote{http://heasarc.gsfc.nasa.gov/docs/suzaku/analysis/abc/} to reprocess and check the data, preliminary light curves and spectra were extracted.  XSPEC 12 was used to fit the spectrum from Suzaku \cite{Arnaud}.  A broken power law modified by interstellar absorption fit the data well, and the result was unfolded and deabsorbed to derive the intrinsic X-ray spectrum.  

In addition, we were able to generate a light curve and spectrum for the VHE $\gamma$-ray data from VERITAS.  The VHE $\gamma$-ray data was fit with a simple power law of the form:

\begin{equation}
\frac{\mathrm{d}N}{\mathrm{d}E} =
 F_0 \cdot 10^{-12} \cdot
 \left(\frac{E}{1\,\mathrm{TeV}}\right)^{-\alpha} \cdot
  \frac{\mathrm{photons}}{\mathrm{TeV}~\mathrm{cm}^{2}~\mathrm{s}}
\end{equation}
	
\noindent Finding best-fit parameters of $\alpha = 2.61\pm0.15$ and $F_0=5.62\pm0.83$.  Uncertainties are statistical only.

The resulting light curves are shown in Figure \ref{lc}.  In the VHE $\gamma$-ray band, the source appears to be in a very quiet state.  For a longer term study of the source's behavior in the VHE $\gamma$-ray band, see \cite{Huang}.  There was no strong variability detected by VERITAS or the HXD on Suzaku, and moderate variability is observed in the XIS light curve.    

\begin{figure}[t]
\centering
\includegraphics[width=80mm]{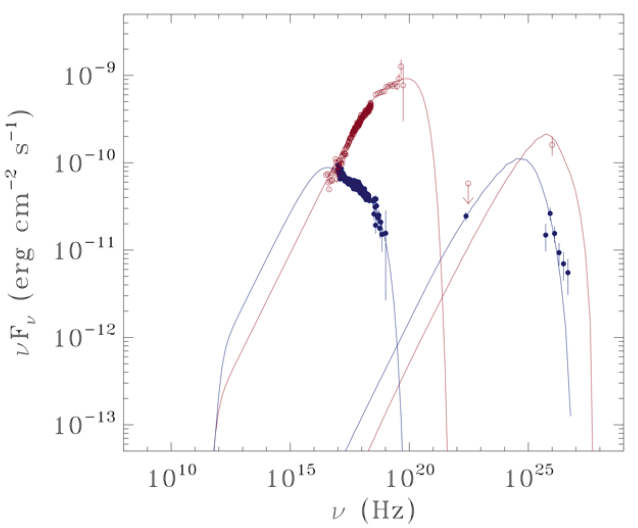}
\caption{Preliminary SED with 2009 data from Suzaku HXD and XIS, Fermi and VERITAS (blue) as well as archival data from Bepposax, EGRET and Whipple (red).} \label{sed}
\end{figure}

The spectral results from the 2009 observatins during the quiescent state as well as archival data from \cite{Pian} and \cite{Catanese} are shown in Figure \ref{sed}.

	The SED data were matched to a simple Synchrotron Self-Compton code written by H. Krawczynski\footnote{http://jelley.wustl.edu/multiwave/spectrum/?code}.  Here it  assumes a spherical blob of radius $1\times10^{15}$ cm, filled with a homogenous non-thermal electron population and uniform magnetic field. The electron population, with $\gamma_{min}$ = 1 where $\gamma$ is the Lorentz factor of the electrons, is described by a broken power law.  
	
	This simple model matches the data well, with the predominant change between the two flux states being the spectrum of injected electrons.  The magnetic field also shifts slighty.  Parameters for the models for the 1997 flaring state and the 2009 quiescent state are given in Table \ref{sscmodel}

\begin{table}[t]
\begin{center}
\caption{SED Model Parameters}
\begin{tabular}{|l|c|c|c|}
\hline \textbf{Model Component} & \textbf{1997 Model} & \textbf{2009 Model} 
\\\hline
Doppler Factor	& 15	&15\\
Magnetic Field (G)&	0.6&	0.95\\
$\gamma_{max}$&	3.00$\times10^{6}$	&3.50$\times10^{5}$\\
$\gamma_{break}$&	3.00$\times10^{5}$&	2.00$\times10^{4}$\\
Index 1 ($\gamma_{min} - \gamma_{break}$)&	1.9	&1.9\\
Index 2 ($\gamma_{break} - \gamma_{max}$)&	2.6	&3.4\\
normalization	&0.8	&0.8\\
\hline
\end{tabular}
\label{sscmodel}
\end{center}
\end{table}

\subsection{Discussion}
This data set provides a high quality sampling of the quiescent state SED of Markarian 501, and it allows detailed comparisons over a broad energy range to be made with the flaring state observed in 1997.
From the X-ray data points alone, the synchrotron emission must peak above 227 keV ($5.49 \times 10^{19}$ Hz) during the flaring state and below 0.4 keV ($9.90 \times 10^{16}$ Hz) during the quiescent state.  The SED model fits imply that the difference must be even greater.  This means that with a change in flux of around one order of magnitude for the quiescent and flaring states used in this study, the synchrotron peak of the SED shifts in frequency by at least 3 orders of magnitude.  

The predominant change in the SED models between the flaring and quiescent state is the spectrum of the electron population, implying that this may be the primary explanation for the dramatic shift in the peak frequency.  The preliminary SSC model appears to indicate that during the flaring state electrons are accelerated up to energies an order of magnitude higher than those in the quiescent state.  However, this match is not unique due to the number of free parameters.  An update of this study, potentially including more data at VHE $\gamma$-ray energies and an improved spectrum from the Fermi Gamma-ray Space Telescope is forthcoming.

\bigskip 
\begin{acknowledgments}
This work was supported in part by NASA through grants NNX08AZ98G and
NNX08AX53G.  We would like to thank Elena Pian for providing the Bepposax data.  The VERITAS research is supported by grants from the U.S. Department of Energy, the U.S. National Science Foundation and the Smithsonian Institution, by NSERC in Canada, by Science Foundation Ireland and by STFC in the UK.  We acknowledge the excel- lentwork of the technical support staff at the FLWO and the collaborating institutions in the construction and operation of the instrument.
\end{acknowledgments}

\bigskip 

\end{document}